\documentclass[10pt]{iopart}

\usepackage{iopams,color,graphicx}

\graphicspath{{eps/}}

\newcommand{\SLE}{{\rm SLE}}

\begin{document}

\title{Discrete holomorphic parafermions in the Ashkin-Teller model and SLE}

\author{Yacine Ikhlef$^1$ and Mohammad Ali Rajabpour$^2$}

\address{$^1$ Section Math\'ematiques, Universit\'e de Gen\`eve
  2-4 rue du Li\`evre, CP 64, 1211 Gen\`eve 4, Switzerland}
  
\address{$^2$ SISSA and INFN, Sezione di Trieste,
  via Bonomea 265, 34136 Trieste, Italy}

\eads{\mailto{y.ikhlef1@physics.ox.ac.uk}, \mailto{rajabpour@sissa.it}}

\maketitle

\begin{abstract}
  We find discrete holomorphic parafermions of the Ashkin-Teller model on the critical line,
  by mapping appropriate interfaces of the model to the ${\rm O}(n=1)$ model. We give support to
  the conjecture that the curve created by the insertion of parafermionic operators at two
  points on the boundary is ${\rm SLE}(4,\rho,\rho)$, where $\rho$ varies along the critical line.
\end{abstract}


\section{Introduction}
\label{sec:introduction}

The discovery of the Schramm-Loewner evolution (SLE) by Schramm~\cite{schramm} opened
a rigorous way to study conformally invariant systems. Some well-known results from 
physics literature were proved~\cite{smirnov1,smirnov2,smirnov3}, new formulas were
also discovered~\cite{SLE-reviews}.
Notably, conformal invariance was proved for percolation
clusters~\cite{smirnov1} and for Ising Fortuin-Kasteleyn and spin clusters~\cite{smirnov2,smirnov3}.
One of the key steps in these proofs is to find an appropriate discrete holomorphic parafermion
for the critical interface.
Relating the holomorphic parafermion to a specific property of the critical curve, one
can conjecture and then prove that the continuum limit of the interface is SLE. Different
methods of finding discrete holomorphic parafermions for statistical models
and their relation to integrability were investigated by Cardy and
collaborators in a series of papers \cite{CR,RC,IC}. In some cases \cite{CR, IC} the relation
of the parafermion to a critical interface is known, and so there are conjectures on the
SLE corresponding to the continuum limit of the interfaces, but in other cases
\cite{RC} the corresponding interfaces are not known and so there is no known SLE. 

One of the interesting models where neither discrete holomorphic parafermions nor corresponding SLE
were known up to now is the Ashkin-Teller (AT) model~\cite{Ashkin,wu_and_lin}
(the phase diagram is described in~\cite{Kadanoff_and_Brown,baxter,Kohmoto_den_nijs}). 
From the Conformal Field Theory (CFT) point of view, the AT model is interesting because it has a
critical line with constant central charge $c=1$ and changing critical exponents.
Although it is possible to define different kinds of critical interfaces for the AT model
\cite{fractal_interfaces,ps}, it is shown numerically~\cite{fractal_interfaces} that some
of the natural possibilities are not related to simple SLE's. Recently, in~\cite{LR}, the AT model
was studied on iso-radial graphs, the critical surfaces on generic iso-radial graphs
were found, and discrete holomorphic parafermions were defined at some particular points of
the critical line, by using algebraic relations between spin and
disorder variables. In the present paper, we introduce discrete holomorphic parafermions
all over the critical line, without using these relations directly. The idea is to map some
particular interfaces to the ${\rm O}(n=1)$ model on the square lattice, and then to exploit some of
the results of~\cite{IC}.

Our main conjecture is based on the continuum Gaussian theory for the Solid-On-Solid (SOS) model
associated to the AT model, and on the known relation between this theory and
$\SLE_4$~\cite{Schramm-Sheffield}: in the upper half-plane, we expect
the curve created by the insertion of a parafermion at the origin and at infinity to have
the statistics of $\SLE(4,\sqrt{g}-1,\sqrt{g}-1)$, where $g$ is the coupling constant of the Gaussian theory
(see Section~\ref{sec:SLE} for details on the definition of $g$).

The structure of the paper is as follows. In Section~2, we recall the definition of the AT model
on the square lattice and its mapping to the staggered six-vertex model. In Section~3, the six-vertex
model is mapped to the ${\rm O}(n=1)$ model, and, using this mapping, we find a lattice holomorphic
parafermion for the AT model all over the critical integrable surface. In Section~4, we formulate
conjectures on the relation of our interfaces to SLE. 
Section~5 is dedicated to bulk critical exponents in the ${\rm O}(n=1)$ loop model
associated to the AT model. The results are checked numerically by transfer-matrix diagonalisation.

\section{The Ashkin-Teller model}
\label{sec:AT-model}

\subsection{Definition and graphical expansion}

\begin{figure}
  \begin{center}
    \includegraphics[scale=1]{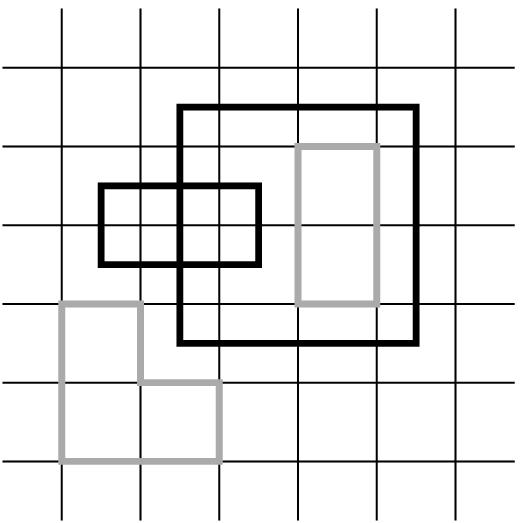}
    \caption{An example configuration in the graph expansion~\eref{eq:Z-cluster}. 
      The spin variables live on the vertices of the underlying square lattice.
      Domain walls for the $\tau'_j$ spins are depicted in thick, black lines,
      and high-temperature polygons for the $\sigma_j$ variables, in thick, grey lines.
      Crossings between the black and grey clusters are forbidden.}
    \label{fig:clusters}
  \end{center}
\end{figure}

The Ashkin-Teller model can be defined on any graph, but, for clarity, we will
restrict to the square lattice in this paper. At each vertex $j$ of the lattice,
we define two spin variables $\sigma_j$ and $\tau_j$, which can take individually
the values $\pm 1$. A spin configuration $\{ \sigma_j, \tau_j \}$ gets the Boltzmann weight:
\begin{equation}
  \fl \qquad
  \prod_{\langle i,j \rangle} W(i,j) \,, \quad
  \hbox{where} \quad W(i,j) = \exp \left[
    \beta_{ij} (\sigma_i \sigma_j + \tau_i \tau_j)
    + \alpha_{ij} \ \sigma_i \sigma_j \tau_i \tau_j
  \right] \,,
\end{equation}
and $\alpha_{ij},\beta_{ij} = \alpha_x,\beta_x$ (resp. $ \alpha_y,\beta_y$) if $\langle i,j \rangle$
is a horizontal (resp. vertical) bond. 
We denote by $\langle \dots \rangle$ the averaging with respect to the normalized 
Boltzmann weights $W(i,j)/Z$, where
\begin{equation}
  \fl \qquad
  Z = \sum_{\{ \sigma_j, \tau_j \}} \prod_{\langle i,j \rangle} W(i,j) \,.
\end{equation}
Let us recall the graphical expansion of this partition function~\cite{nienhuis2,Saleur}.
Introducing the change of variables $(\sigma_j,\tau_j) \to (\sigma_j,\tau'_j)$,
where $\tau'_j=\sigma_j \tau_j$, the edge interaction can be written as:
\begin{equation}
  \fl \qquad
  W(i,j) = e^{\alpha_{ij} \tau'_i \tau'_j} \left[
    \cosh \beta_{ij} (1+\tau'_i \tau'_j)
    + \sigma_i \sigma_j \ \sinh \beta_{ij} (1+\tau'_i \tau'_j)
  \right] \,.
\end{equation}
First, we fix the values of the $\tau'_j$ spins, which define a domain-wall (DW) configuration
on the edges of the dual lattice. Let $\langle i,j \rangle$ be an edge of the original lattice.
If $\langle i,j \rangle$ is crossed by a DW, then it gets a weight $e^{-\alpha_{ij}}$. If not, then
it gets the weight $e^{\alpha_{ij}} [\cosh 2\beta_{ij} + \sigma_i \sigma_j \ \sinh 2\beta_{ij}]$. We depict the first 
term as an empty edge, and the second one as an occupied edge.
Thus the partition function reads
\begin{eqnarray}
  \fl \qquad
  Z = {\rm const} \times &\sum_{G|G'}&
  (\tanh 2\beta_x)^{\ell_x(G)}
  (\tanh 2\beta_y)^{\ell_y(G)} \nonumber \\
  && \times (e^{2\alpha_y} \cosh 2\beta_y)^{-\ell_x(G')}
  (e^{2\alpha_x} \cosh 2\beta_x)^{-\ell_y(G')} \,,
  \label{eq:Z-cluster}
\end{eqnarray}
where $G$ (resp. $G'$) is a subgraph of the original (resp. dual) lattice,
all vertices in $G$ and $G'$ must have even degree,
$G$ and $G'$ must not intersect each other, and $\ell_x(G), \ell_y(G)$
are the numbers of horizontal and vertical bonds of $G$. An example configuration is
shown in Figure~\ref{fig:clusters}.

\subsection{Mapping to the six-vertex model}

\begin{figure}
  \begin{center}
    \includegraphics[scale=1]{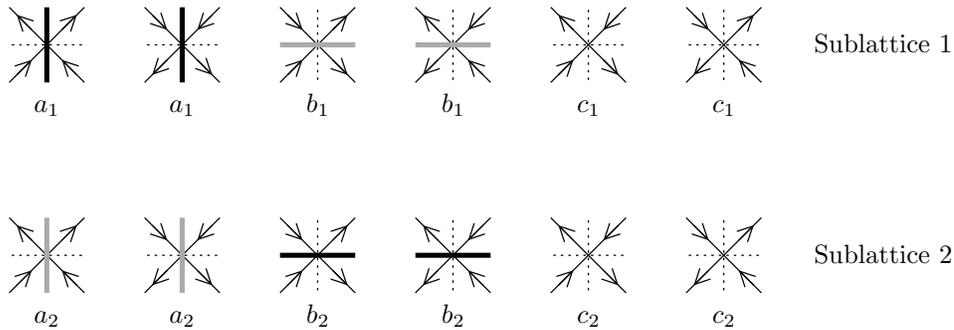}
    \caption{Mapping from the graph expansion of the AT model to the six-vertex model.
    The colour of the possible graph bond depends on the sublattice of the medial lattice.}
    \label{fig:AT-6V}
  \end{center}
\end{figure}

One interest of this graphical expansion is that it maps to
the six-vertex model on the medial lattice\footnote{
  An earlier construction, described in~\cite{baxter}, relates the AT model to
  a staggered eight-vertex model, without using the $\tau'$ variables. However,
  in the present paper, we focus on the mapping to the six-vertex model because
  the latter has a known discrete holomorphic parafermion.
}. The correspondence is given in
Figure~\ref{fig:AT-6V}.
The Ashkin-Teller model is critical when the six-vertex model is not staggered.
The weights of the six-vertex model with parameter $\Delta = (a^2+b^2-c^2)/(2ab)=-\cos \lambda$
can be parameterised as
\begin{equation} \label{eq:weights-6V}
  \fl \qquad
  a,b,c = \sin(\lambda-u), \sin u, \sin \lambda \,.
\end{equation}
This corresponds to the weights of the AT model:
\begin{eqnarray}
  \fl \qquad
  \tanh 2\beta_x = \frac{\sin u}{\sin \lambda}
  \,, &\qquad& \frac{e^{-2\alpha_x}}{\cosh 2\beta_x} = \frac{\sin (\lambda-u)}{\sin \lambda}
  \label{eq:critical1} \\
  \fl \qquad
  \tanh 2\beta_y = \frac{\sin (\lambda-u)}{\sin \lambda}
  \,, &\qquad& \frac{e^{-2\alpha_y}}{\cosh 2\beta_y} = \frac{\sin u}{\sin \lambda} \,.
  \label{eq:critical2} 
\end{eqnarray}
The variable $u$ is a spectral parameter, whereas $\lambda$ determines the universality class.
The isotropic point is at $u=\lambda/2$.
Some special values are: $\lambda_{\rm P}=0$, $\lambda_{\rm FZ}=\frac{\pi}{4}$, $\lambda_{\rm I} = \frac{\pi}{2}$ and
$\lambda=\frac{3\pi}{4}$, corresponding to the four-state Potts, $\mathbb{Z}_4$ Fateev-Zamolodchikov \cite{ZF},
Ising$\times$Ising and XY models. When $\lambda$ is varied, the central charge remains constant $c=1$,
but the critical exponents change. For example, the correlation exponent is
\begin{equation}
  \fl \qquad
  \nu(\lambda) = \frac{2\pi-2\lambda}{3\pi-4\lambda} \,.
\end{equation}

More generally, it was shown in Ref~\cite{LR} that the critical weights of the AT models
on any Baxter lattice are given by~\eref{eq:critical1}--\eref{eq:critical2}, with the spectral
parameter $u$ related to the angle of $\theta$ the rhombic faces by $\theta=\pi  u/\lambda$.
In the next Section, we will describe a discrete holomorphic parafermion
in the critical square-lattice AT model. Note that this parafermion is also present for the AT
model on a Baxter lattice, at the critical value of the Boltzmann weights.

\begin{figure}
  \begin{center}
    \includegraphics[scale=1]{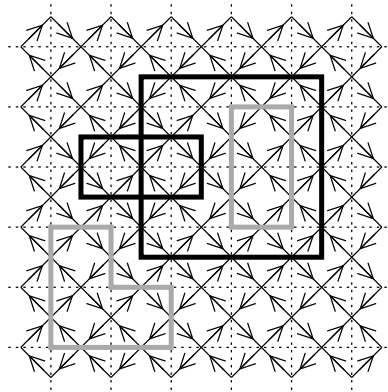}
    \caption{Arrow configuration corresponding to the graph configuration shown in
      Figure~\ref{fig:clusters}. The arrows of six-vertex model live on the edges of
      the medial lattice.}
    \label{fig:clusters-6V}
  \end{center}
\end{figure}

\section{Discrete holomorphic parafermion}
\label{sec:parafermions}

\begin{figure}
  \begin{center}
    \includegraphics[scale=0.8]{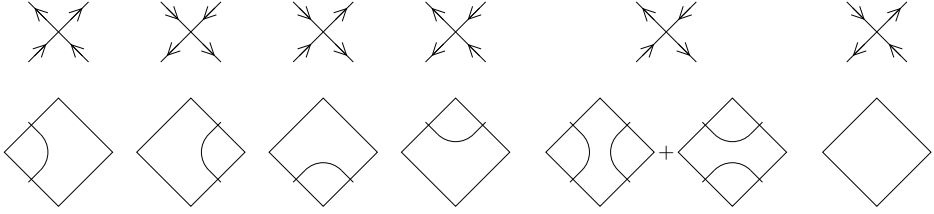}
    \caption{Mapping between the six-vertex model and an ${\rm O}(n=1)$ loop model.
      The correspondence between the arrow and loop vertices shown here is valid
      on the sublattice 1 of the medial lattice. On the sublattice 2, the same
      correspondence holds, with all arrows reversed.}
    \label{fig:6V-On}
  \end{center}
\end{figure}

In~\cite{LR}, using the spin variables of the AT model, discrete holomorphic parafermions
were found at some particular points of the critical line, namely at the four-state Potts,
FZ, Ising and XY points. In this Section, we will exhibit {\it another} discrete
holomorphic parafermion all along the critical line of the AT model. This parafermion is
defined in the loop formulation of the AT model.
The key idea is to use the chain of mappings:
\begin{equation*}
  \fl \qquad
  \hbox{Ashkin-Teller} \longrightarrow \hbox{Six-vertex}
  \longrightarrow \hbox{${\rm O}(n=1)$} \,.
\end{equation*}
The first mapping is described in Section~\ref{sec:AT-model}, and the second one is given,
for example, in~\cite{IC}. For completeness, we recall it in Figure~\ref{fig:6V-On}.
The resulting loop model has seven possible vertices, and each closed loop has
a Boltzmann weight $n=1$. We will denote this model the ${\rm O}(n=1)$ model for short.

It was shown in~\cite{IC} that the ${\rm O}(n=1)$ model, with weights given
by~\eref{eq:weights-6V}, possesses a discrete holomorphic parafermion $\psi(z)$,
associated to the insertion of one path at point $z$. The spin $s$ of this parafermion
is related to the parameter $\lambda$ by:
\begin{equation} \label{eq:s-lambda}
  \fl \qquad
  s = 1 - \frac{\lambda}{\pi} \,.
\end{equation}

In terms of the AT model, $\psi$ is defined on the edges of the
covering lattice (the union of the original and dual lattices).
If $z$ denotes an edge of this lattice,
the operator $\psi(z)$ inserts a $\sigma$ variable and a $\tau'$
DW at the ends of the edge $z$. The two-point function $\langle \psi(z_1) \psi(z_2) \rangle$ also contains
a non-local phase factor given by the simultaneous winding of the $\tau'$ DW and $\sigma$
cluster between $z_1$ and $z_2$. We might decompose symbolically $\psi$ into three factors:
\begin{equation}
  \psi = e^{i\theta} \times \sigma \times \mu_{\tau'} \,,
\end{equation}
where $\theta$ is a contribution to the winding angle (see above) and $\mu_{\tau'}$ creates a
$\tau'$ domain wall.

\begin{figure}
  \begin{center}
    \includegraphics[scale=1]{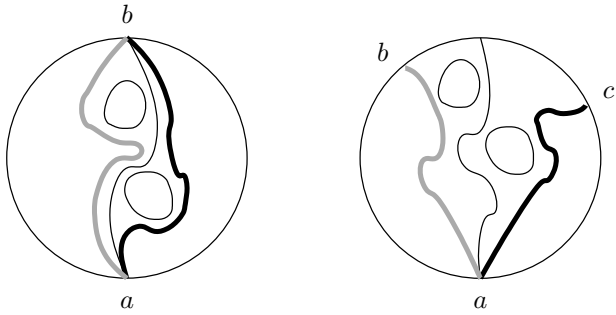}
    \caption{Random curves associated to the parafermion $\psi$, defined on a
      domain $\Omega$ with boundary conditions BC$_1$ (left) and BC$_2$ (right).
      In both cases, closed loops may be present around $\gamma$.}
    \label{fig:domain}
  \end{center}
\end{figure}

Let us interpret the effect of $\psi$ on the AT model, when inserted at two boundary
points $a$ and $b$. In this situation, there must be, in the ${\rm O}(n=1)$ model,
a path $\gamma$ going from $a$ to $b$ (see Figure~\ref{fig:domain}).
The occupied area adjacent to $\gamma$ is bounded on
one side by a $\tau'$ domain wall, and on the other side by a $\sigma$ high-temperature
cluster. So $\langle \psi(a) \psi(b) \rangle$ corresponds to the two-point function
$\langle \sigma(a) \sigma(b) \rangle_{1,ab}$,
where $\langle \dots \rangle_{1,ab}$ denotes the averaging with Boltzmann weights $W(i,j)/Z$ and the following
boundary conditions (denoted BC$_1$): free boundary conditions for $\sigma$ and $\tau'=1$ on
one arc $(ab)$ of the boundary, $\tau'=-1$ on the other arc.

We can also think of boundary conditions which allow the path $\gamma$ to end anywhere in a
given interval of the boundary.
Let $a,b,c$ be three marked points on the boundary.
We consider the two-point function $\langle \sigma(a) \sigma(b) \rangle_{2,abc}$,
where $\langle \dots \rangle_{2,abc}$ is the Boltzmann average
with the following boundary conditions: free BC for $\sigma$, $\tau'=+1$ on $(ab)$ and $(bc)$,
$\tau'=-1$ on $(ac)$. In the ${\rm O}(n=1)$ model, this forces a path $\gamma$ to go from 
$a$ to a point on $(bc)$, as shown in Figure~\ref{fig:domain}.
We call these boundary conditions BC$_2$.
If $\mu_{\tau'}$ is viewed as a boundary-condition changing operator, the correlation
function $\langle \sigma(a) \sigma(b) \rangle_2$ can be written as a three-point function, reflecting
explicitly the role of point $c$:
\begin{equation}
  \langle \sigma(a) \sigma(b) \rangle_{2,abc}
  = \langle \psi(a) \sigma(b) \mu_{\tau'}(c) \rangle
\end{equation}

\section{Relation to SLE}
\label{sec:SLE}

The six-vertex model can be transformed into an SOS model by introducing height variables
$\varphi$ on the faces of the medial lattice, so that two neighbouring $\varphi$'s differ
by $\pm \pi/2$, the highest being on the left of each arrow.
The SOS model renormalises to a Gaussian theory with action
\begin{equation} \label{eq:action}
  \fl \qquad
  A = \frac{g}{4\pi} \int |\nabla \varphi|^2 {\rm d}^2x \,,
  \qquad g = \frac{4(\pi-\lambda)}{\pi} \,.
\end{equation}

When the AT model is defined on a system with one or more periodic directions
({\it e.g.}, a torus or a cylinder), the height $\varphi$ is only well-defined
up to the identifications $\varphi \equiv \varphi+\pi, \varphi \equiv -\varphi$. The first corresponds to
a height defect around one direction of the system, while the second is a `twist' induced
by an odd number of DW's winding around the system~\cite{Saleur}. The theory with
action~\eref{eq:action} and the above identifications is called the
$\mathbb{Z}_2$-orbifold of the compact boson. However, in this Section, we will be dealing only with
simply-connected domains, so the SOS configurations are always well-defined without
any identification, and the continuum limit is simply the Gaussian model~\eref{eq:action}.

In the case of boundary conditions BC$_1$, the presence of the path $\gamma$ induces a 
height gap at $a$ and $b$. So we expect the SOS model to be a free field with Dirichlet BC,
and boundary values $-\varphi_1$ on one arc, $\varphi_2$ on the other arc.
Since the DW and high-T cluster play the same role at the critical point, we must have $\varphi_1=\varphi_2$.
Schramm and Sheffield have shown~\cite{Schramm-Sheffield} that the contour line
in such a model is $\SLE(4,\rho_1, \rho_2)$, where $\rho_1=\varphi_1/\varphi^*-1,
\rho_2=\varphi_2/\varphi^*-1$, and $\varphi^*$ is a universal constant.
Now, we use the results of ~\cite{Affleck,Cardy2004}: when a consistent normalisation
is chosen for $\varphi$, we have $\varphi^* = \pi/\sqrt{4g}$ and the operator inserting the
height gap $\delta\varphi$ has conformal dimension
\begin{equation*}
  \fl \qquad
  h= g \left( \frac{\delta\varphi}{2\pi} \right)^2 \,.
\end{equation*}
Comparing to the
spin~\eref{eq:s-lambda} of the parafermion $s=g/4$, we obtain $\varphi_1=\varphi_2 = \pi/2$,
and so we conjecture that the curve $\gamma$ has the statistics of
$\SLE(4,\sqrt{g}-1,\sqrt{g}-1)$.


\section{Critical exponents and fractal dimension}
\label{sec:exponents}

In this Section, we derive the bulk exponents for watermelon correlation functions
in the ${\rm O}(n=1)$ model, and give numerical results on some of these exponents.
The ${\rm O}(n=1)$ model is a peculiar loop model, where the central charge is fixed,
but the critical exponents vary along the critical line.

To calculate bulk exponents, it is easiest to consider the loop model on a cylinder of
circumference $L$ sites ($L$ even). In this setting, the model is described by the
$\mathbb{Z}_2$-orbifold theory, and part of the conformal spectrum~\cite{Saleur} is given by
`electromagnetic' exponents
\begin{equation}
  \fl \qquad
  X_{em} = \frac{e^2}{2g} + \frac{g m^2}{2} \,.
\end{equation}

In the transfer-matrix formalism, the $k$-leg watermelon exponent
corresponds to the dominant eigenvalue in the sector with $k$ strands propagating along the
cylinder. For a given $k$, there are several exponents, according to the parity of the
sites where the $k$ strands sit. Indeed, because of the staggering in the
6V/${\rm O}(n=1)$ correspondence, strands sitting on an even (resp. odd) edge are oriented
positively (resp. negatively). Each strand contributes a height defect of magnetic charge $m=\pm 1/2$,
so if we write $k=k_1+k_2$, where $k_1,k_2$ are the numbers
of strands sitting on even and odd edges, the total magnetic charge is $m=(k_1-k_2)/2$.
Hence, when $k_1 \neq k_2$, we get the exponent:
\begin{equation}
  \fl \qquad
  X_{k_1,k_2} = \frac{g (k_1-k_2)^2}{8} \,, \qquad k_1 \neq k_2 \,.
\end{equation}
In the special case $k_1=k_2$, the total magnetic charge vanishes. Our numerical results
allow us to conjecture to exponent:
\begin{equation}
  \fl \qquad
  X_{\ell,\ell} = \frac{\ell^2}{2} \,.
\end{equation}

The fractal dimension of the path $\gamma$ is related to a two-leg watermelon exponent:
$d_f=2-X$. For $k=2$, there are two choices: $X_{2,0}=g/2$ and $X_{1,1}=1/2$. The choice $X_{1,1}$ gives
a fractal dimension $d_f=3/2$, which is the correct value for $\SLE_4$. Note that
the direct numerical calculations of Picco and Santachiara~\cite{ps2} on the AT model show
that the fractal dimension of the boundaries of $\tau'$ clusters is also $3/2$.

\begin{figure}
  \begin{center}
    \includegraphics[scale=1]{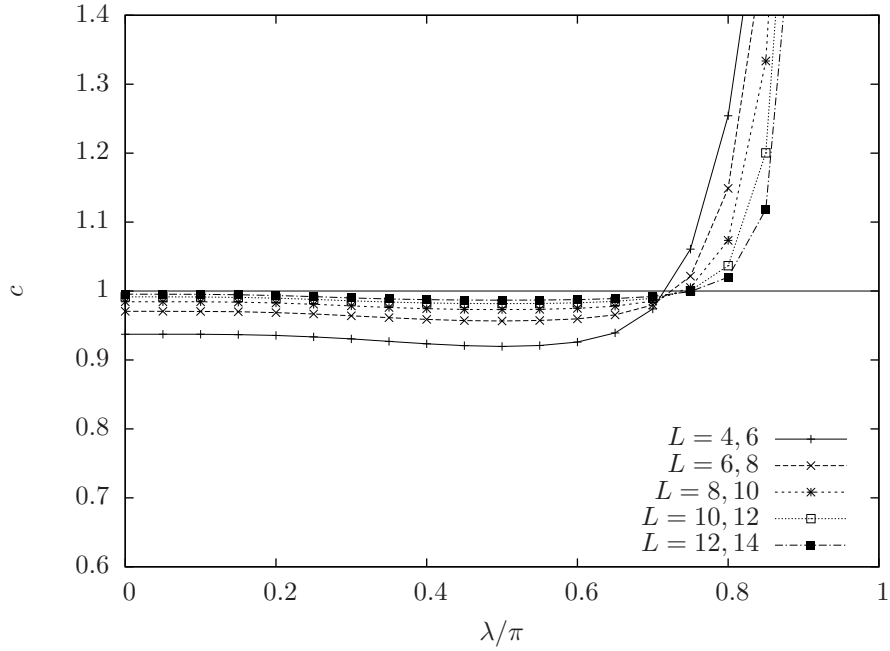}
    \caption{Central charge of the loop model. The expected value is $c=1$
      on the whole critical line.}
    \label{fig:cc}
  \end{center}
\end{figure}

\begin{figure}
  \begin{center}
    \includegraphics[scale=1]{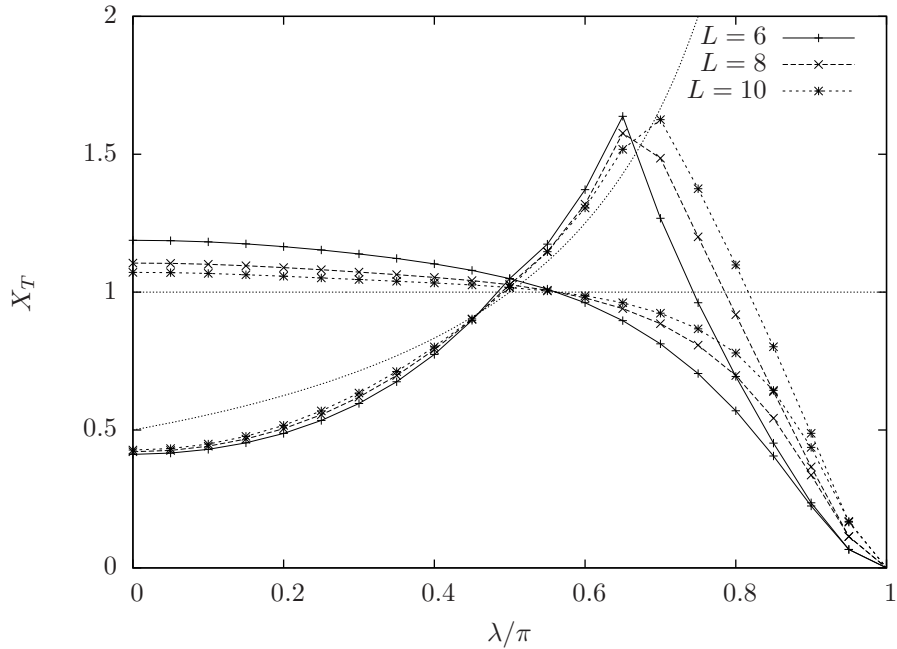}
    \caption{First and third thermal exponents of the loop model.
      There is a level crossing between the exponents
      $X=2-1/\nu(\lambda)$ and $X=1$.}
    \label{fig:xt}
  \end{center}
\end{figure}

\begin{figure}
  \begin{center}
    \includegraphics[scale=1]{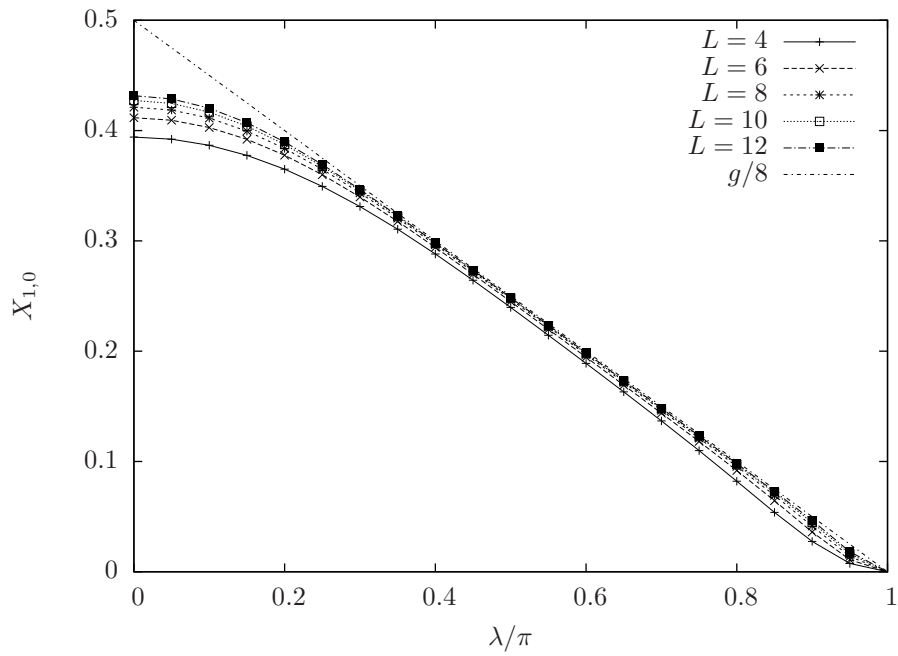}
    \caption{One-leg watermelon exponent of the loop model.}
    \label{fig:x1}
  \end{center}
\end{figure}

\begin{figure}
  \begin{center}
    \includegraphics[scale=1]{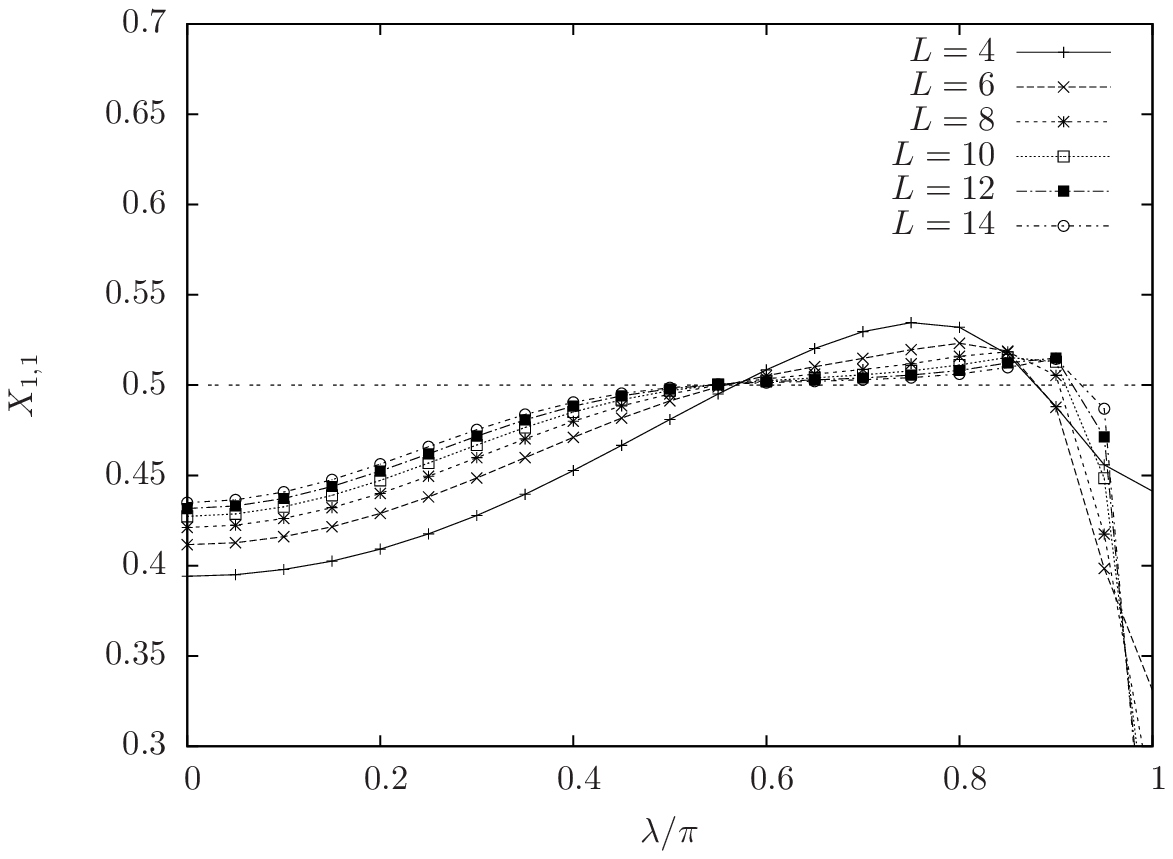}
    \caption{Two-leg watermelon exponent $X_{1,1}$ of the loop model.}
    \label{fig:x2}
  \end{center}
\end{figure}

\begin{figure}
  \begin{center}
    \includegraphics[scale=1]{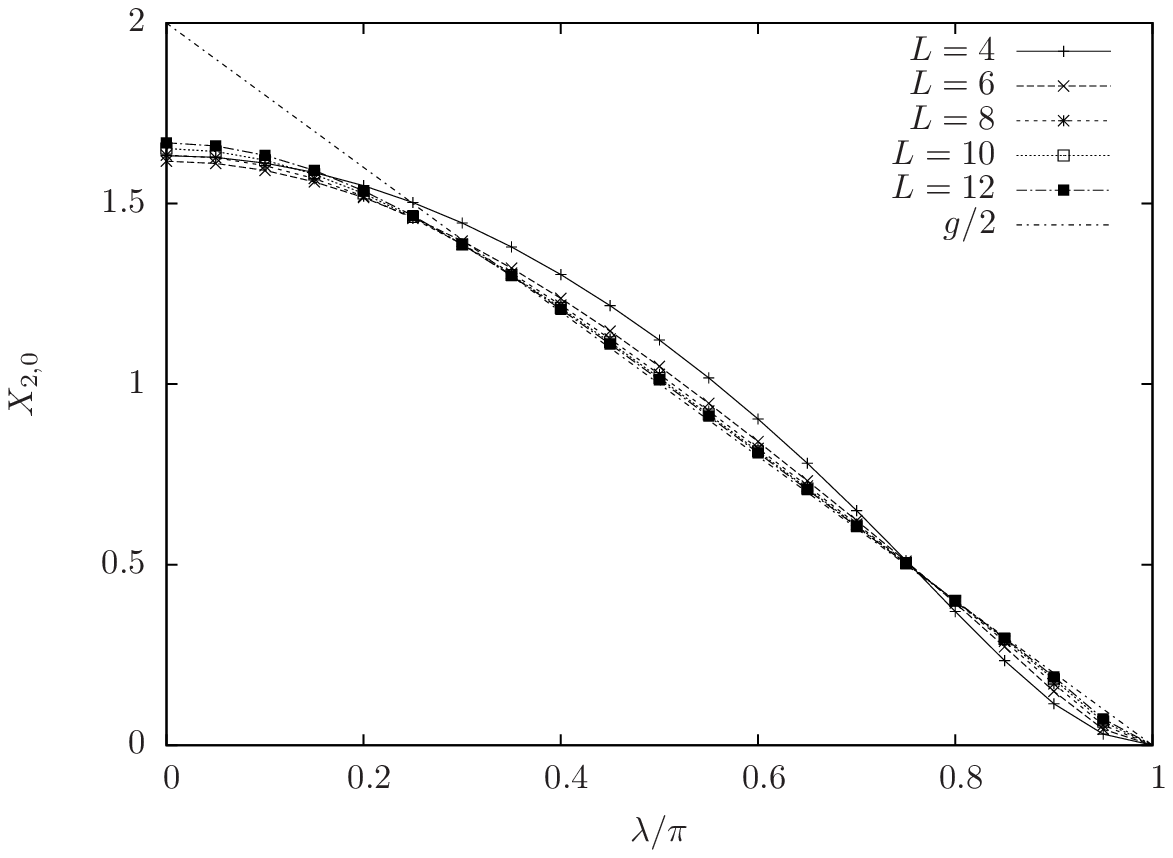}
    \caption{Two-leg watermelon exponent $X_{2,0}$ of the loop model.}
    \label{fig:x2p}
  \end{center}
\end{figure}

\begin{figure}
  \begin{center}
    \includegraphics[scale=1]{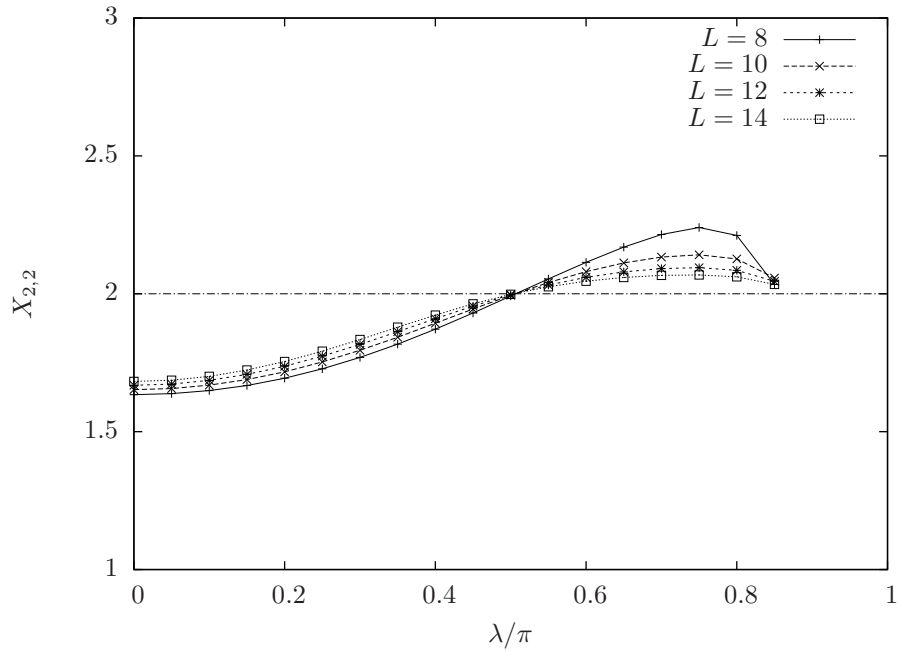}
    \caption{Four-leg watermelon exponent $X_{2,2}$ of the loop model.}
    \label{fig:x4}
  \end{center}
\end{figure}

\begin{figure}
  \begin{center}
    \includegraphics[scale=1]{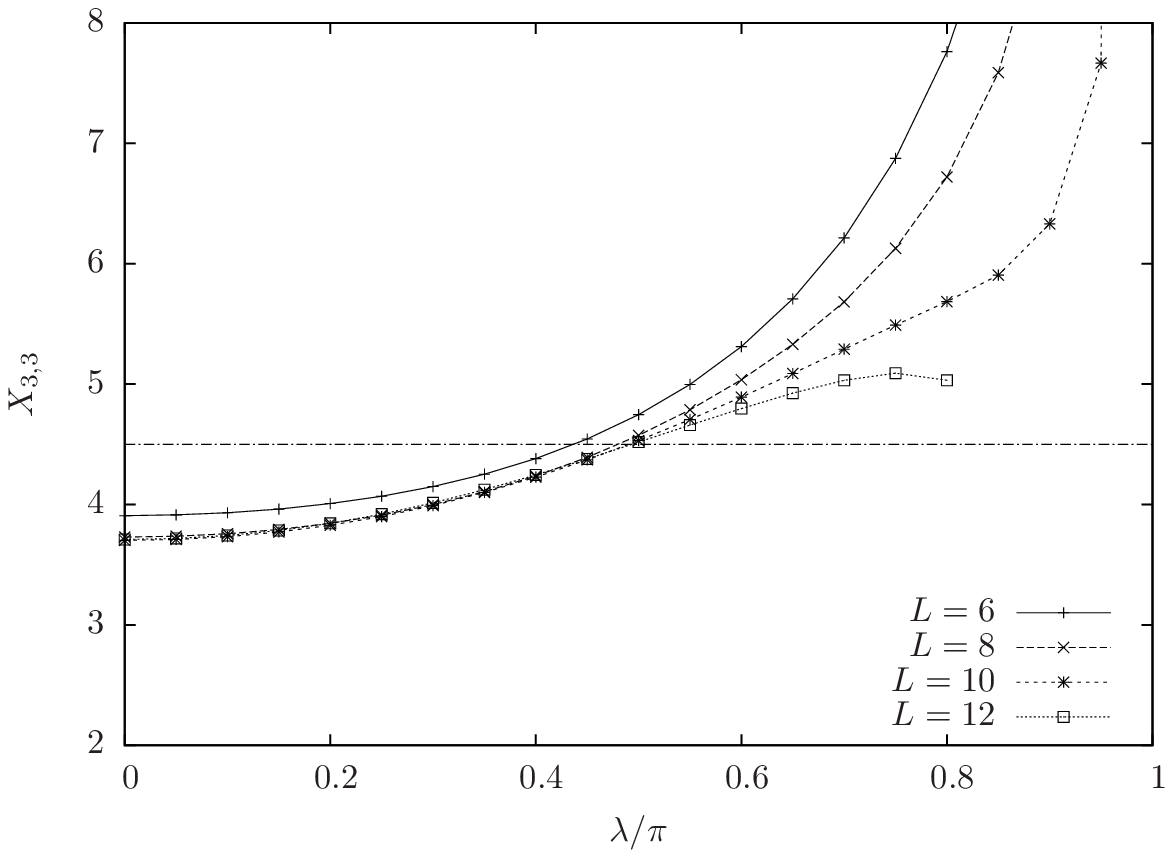}
    \caption{Six-leg watermelon exponent $X_{3,3}$ of the loop model.}
    \label{fig:x6}
  \end{center}
\end{figure}

The central charge and the exponents of the loop model are extracted using the finite-size
formulae for the eigenvalues $\Lambda_j$ of the transfer matrix~\cite{Cardy-FS}:
\begin{equation}
  \fl \qquad
  - \log \Lambda_j \simeq Lf_\infty
  + \frac{2\pi}{L} \left(-\frac{c}{12}+X_j \right) \,.
\end{equation}
In Figures~\ref{fig:cc}--\ref{fig:x6}, we depicted numerical data for the central charge, the thermal
exponent and some watermelon exponents of the ${\rm O}(n=1)$ loop model.

\section{Conclusion}
\label{sec:Conclusion}

In this paper we first found a discrete holomorphic parafermion for AT model on the whole
critical line. There may exist other discrete holomorphic parafermions in this model, but
the one we describe is defined in terms of the ${\rm O}(n=1)$ loop model, which enables us to
relate it to the SLE model. Our conjecture is compatible with our numerical calculations
on the $O(n=1)$ model. 
Of course, more precise numerical calculation is needed to confirm our conjecture, such as
the left-right Schramm's formula. Unfortunately, no Schramm's formula for $\SLE(\kappa,\rho_1,\rho_2)$
is known, but for $\kappa=4$ it could be tractable, thanks to the relation with the Gaussian Free Field.
One can also think of Monte-Carlo simulation as it was done
for the Ising model~\cite{dipolar}, and martingale arguments~\cite{Kalle-Clement}.

At the point $\lambda=\pi/2$, the AT model is equivalent to two decoupled Ising models.
Oshikawa and Affleck have studied the boundary CFT of this model, in relation to the $\mathbb{Z}_2$
orbifold theory~\cite{Affleck}. In particular, our parafermionic observable $\psi$, when
inserted on the boundary, corresponds to a jump in Dirichlet boundary conditions
for the orbifold ($\varphi_0=0 \to \pi$ in the notations of~\cite{Affleck}), together with the
insertion of a $\sigma$ operator. An interesting direction would be to try and extend these results
to the critical line of the AT model, and analyse the null-vector equations of the boundary
operators, to relate them properly to SLE.

\ack{M. A. Rajabpour benefitted from many fruitful discussions with
R. Santachiara, M. Picco, K. Kyt\"{o}l\"{a}, M. Caselle and S. Lottini. M. A. Rajabpour would like
to thank the Section de Math\'ematiques at Geneva University, where this work was achieved,
for its hospitality.}

\end{document}